\documentclass[dvipdfm]{aa}
\usepackage[dvips]{graphicx}
\usepackage{txfonts}
\usepackage{natbib}
\bibpunct{(}{)}{;}{a}{}{,}

\begin{document}
\title{New $\lambda$6~cm and $\lambda$11~cm observations of the supernova remnant CTA 1}
\author{X. H. Sun\inst{1,2}
        \and W. Reich\inst{2}
        \and C. Wang\inst{1}
        \and J. L. Han\inst{1}
        \and P. Reich\inst{2}}
\institute{National Astronomical Observatories, CAS, Jia-20 Datun Road, 
            Chaoyang District, Beijing 100012, China\\
            \email{xhsun@nao.cas.cn}
            \and 
           Max-Planck-Institut f\"{u}r Radioastronomie, Auf dem H\"ugel 69, 53121 Bonn, Germany\\
           \email{wreich@mpifr-bonn.mpg.de}}
\date{Received / Accepted}

\abstract
{}
{We attempt to study spatial variations of the spectrum and rotation 
measures (RMs) of the large diameter, high-latitude supernova remnant (SNR) 
CTA~1.
}
{We conducted new $\lambda$6~cm and $\lambda$11~cm observations of CTA~1 using 
the Urumqi 25-m and Effelsberg 100-m telescopes. Data at other wavelengths were 
included to investigate the spectrum and polarisation properties.}
{We obtained new total intensity and polarisation maps at $\lambda$6~cm and 
$\lambda$11~cm with angular resolutions of $9\farcm5$ and $4\farcm4$, 
respectively. We derived a spectral index of $\alpha=-0.63\pm0.05$ 
($S_\nu\propto\nu^\alpha$) based on the integrated flux densities at 408~MHz, 
1420~MHz, 2639~MHz, and 4800~MHz. The spectral index map calculated from data 
at the four frequencies shows a clear steepening of the spectrum from the 
strong shell emission towards the north-western breakout region with weak 
diffuse emission. The decrease of the spectral index is up to about 
$\Delta\alpha=0.3$. The RM map derived from polarisation data at $\lambda$6~cm 
and $\lambda$11~cm shows a sharp transition between positive RMs in the 
north-eastern and negative RMs in the south-western part of the SNR. We note a 
corresponding RM pattern of extragalactic sources and propose the existence of 
a large-diameter Faraday screen in front of CTA~1, which covers the 
north-eastern part of the SNR. The RM of the Faraday screen is estimated to be 
about +45~rad~m$^{-2}$. A RM structure function of CTA~1 indicates a very 
regular magnetic field within the Faraday screen, which is larger than about 
2.7~$\mu$G in case of 500~pc distance.}
{CTA~1 is a large-diameter shell-type SNR located out of the Galactic plane, 
which makes it an ideal object to study its properties without suffering 
confusion. The rare breakout phenomenon known for CTA~1 is confirmed. We 
identify a Faraday screen partly covering CTA~1 with a regular magnetic field 
in the opposite direction to the interstellar magnetic field. The detection of 
Faraday screens in the Galactic plane is quite common, but is difficult at high 
latitudes where the polarisation angles of weak polarised background emission 
are rotated. RMs from extragalactic sources are needed for this purpose, 
although the number density of extragalactic RMs is still small despite of 
significant observational progress.}


\maketitle

\section{Introduction}
The supernova remnant (SNR) CTA 1 (G119.5+10.2) was discovered by \citet{hr60}. 
Since then extensive observations at radio, optical, and X-ray bands have been 
made. Radio maps at high-angular resolution were observed with the Effelsberg 
100-m telescope at 1720~MHz and 2695~MHz \citep{shs79, ssm81} and with the DRAO 
synthesis array combined with data from the Effelsberg 100-m telescope  
at 408~MHz and 1420~MHz \citep{plmg93,plsr97}. CTA 1 shows a well-defined 
semi-circular shell towards the south-east and weak diffuse emission towards 
the north-west. Optical observations revealed strong [\ion{O}{III}] filaments, 
which generally coincide with the radio shell \citep{fbk+81,fgk83,mppv00}. 
Strong centre-filled X-ray emission was also observed, which consists of 
both a thermal component heated by the reverse shock and a non-thermal component
powered by a then putative pulsar \citep{sss95, ssb+97}. The pulsar together 
with its wind nebula was recently detected by {\it Fermi} \citep{aaa+08}. All 
observations indicate that CTA~1 is a composite SNR in the Sedov phase.

CTA~1 is a text-book example for the breakout phenomenon \citep{plsr97}, 
similar to the SNR VRO 42.05.1 \citep{lprv82}. The blast wave of the supernova 
sweeps the low-density region towards the north-west, where the magnetic field 
is not sufficiently compressed and the limb-brightened shell is not formed. 
Breakout phenomena of SNRs can be used to study the non-uniformity of the 
interstellar medium. Spatial variations in the spectral index could be a 
diagnosis of the particle acceleration mechanisms. The steepening of the 
spectrum towards the diffuse region was qualitatively shown by 
\citet{plsr97} based on 408~MHz and 1420~MHz data. However, observations at 
higher frequencies are needed to obtain a more precise spatial distribution of 
the spectral index. 

CTA~1 was argued to be a site of enhanced interstellar plasma turbulence by 
analysing the structure functions for rotation measures (RMs) of extragalactic 
sources in the line-of-sight through and outside the SNR \citep{sim92}. In 
principle it would be straightforward to compare the RM structure function of 
extragalactic sources directly with that of the SNR. Polarised emission has been 
observed at 2695~MHz \citep{ssm81} and 1720~MHz \citep{shs79}. However, 
measurements at these two bands cannot warrant a reliable RM map for the SNR 
because of depolarisation at 1720~MHz. Polarisation observations at higher 
frequencies such as 4.8~GHz are therefore essential to calculate RMs.
    
CTA~1 resides at $l=119\fdg5$, $b=10\fdg2$ and experiences little obscuration 
by the emission from the plane at that high latitude. This makes it a 
well-suited object to study the spatial variation of the spectral index. In 
this paper we present new $\lambda$6~cm and $\lambda$11~cm observations of 
CTA~1, which is one of the targets of the campaign to map large northern sky 
SNRs at $\lambda$6~cm. This project is part of the Sino-German $\lambda$6~cm 
polarisation survey of the Galactic plane 
\citep{shr+07,grh+10,srh+11,xhr+11,ghr+11}. Results on some large SNRs have 
been already reported, such as G65.2+5.7 \citep{xrfh09}, the Cygnus Loop 
\citep{srh+06}, HB~3 \citep{shg+08}, G156.2+5.7 \citep{xhs+07}, and S~147 
\citep{xfrh08}.

We describe the new $\lambda$6~cm and $\lambda$11~cm observations of CTA~1 as 
well as other data used, and present the maps in Sect.~2. The results and 
discussions on the spectrum of integrated flux densities, maps of spectral 
index and RM are given in Sect.~3. Particularly the detection of a large 
high-latitude Faraday screen is reported. We summarize our results in Sect.~4.

Throughout the paper we adopt the distance to the SNR of $1.4\pm0.3$~kpc 
determined from \ion{H}{I} measurements \citep{plmg93}. 

\section{New observations and other data used}

\subsection{Urumqi $\lambda$6~cm observations}
CTA~1 was observed with the Urumqi 25-m telescope of National Astronomical 
Observatories, Chinese Academy of Sciences, between September and December 
2004. Details on the $\lambda$6~cm receiving system were already given by 
\citet{srh+06,shr+07}. A field of $3\degr\times3\degr$ size, centred at 
${\rm RA}=0^{\rm h}15^{\rm m}$, ${\rm Dec}=72\degr48\arcmin$ (unless otherwise 
noted, all the Equatorial coordinates are in the epoch of J2000.0 hereafter), 
was observed in raster-scans along both Right Ascension and Declination 
directions. All relevant observational parameters are listed in 
Table~\ref{obs}. 

\begin{table}[!htbp]
\caption{Observation parameters at $\lambda$6~cm and $\lambda$11~cm for CTA~1.
\label{obs}}
\begin{tabular}{lll}\hline\hline
Wavelength                     &  $\lambda$6~cm & $\lambda$11~cm \\
Frequency                      &  4.8~GHz       &  2.639~GHz     \\
Bandwidth                      &   600~MHz      &  80~MHz        \\
Resolution                     &   $9\farcm5$   &  $4\farcm4$    \\
$T_{\rm sys}$                  &    22~K        &  17~K          \\ 
$T_{\rm B}$ [K]/$S$ [Jy]       &    0.164       &  2.52          \\
Scan velocity                  &  $2\degr$/min  & $3\degr$/min  \\
Sub-scan separation            &   $3\arcmin$   & $2\arcmin$       \\
Coverages                      &     9          &  4               \\
rms of $I$ ($\sigma_I$)        &    0.5~mK      &  3~mK           \\
rms of $PI$ ($\sigma_{PI}$)      &    0.3~mK      &  1~mK           \\
Primary calibrator             &  3C 286        &  3C 286         \\
~~Flux density                 &   7.5~Jy       &   10.4~Jy       \\
~~Polarisation percentage      &   11\%         &   9.9\%         \\
~~Polarisation angle           &   $33\degr$    &    $33\degr$    \\ 
\hline
\end{tabular}
\end{table}

The data processing procedures were described by e.g. \citet{shr+07} and 
\citet{grh+10}. The raw data from the observations are maps of Stokes $I$, $U$, 
and $Q$ stored in \textsc{NOD2} format \citep{has74}. For each individual map, 
spikes were removed, the baselines were adjusted, and the scanning effects were 
suppressed using the method developed by \citet{sr79}. The edited map was 
further multiplied by a calibration factor determined from 3C~286 to convert 
map-units into units of main-beam brightness temperature. Maps observed in 
orthogonal directions were then added in the Fourier domain \citep{eg88} to 
further eliminate scanning effects and yield the final results. Instrumental 
polarisation was cleaned according to observations of the unpolarised 
calibrator 3C~295 following the method described by \citet{shr+07}.

\begin{figure}[!htbp]
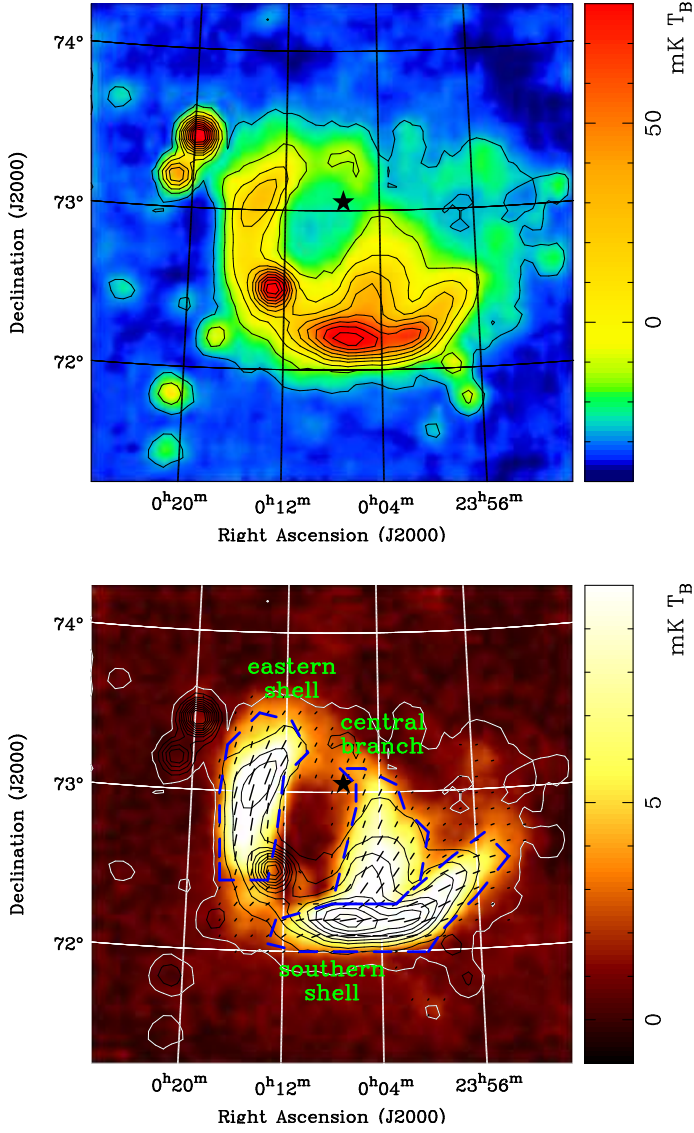

\centering
\resizebox{0.49\textwidth}{!}{\includegraphics[angle=-90]{cta1.i.ps}}\\[5mm]
\resizebox{0.49\textwidth}{!}{\includegraphics[angle=-90]{cta1.ipib.ps}}
\caption{Urumqi $\lambda$6~cm maps. Total intensities are displayed as 
image and contours in the {\it upper panel}. Polarisation intensities are 
displayed as image in the {\it lower panel}, overlaid with bars showing B-vectors 
(i.e. observed E-vectors rotated by $90\degr$) and total intensity contours. 
The lengths of the bars are proportional to polarisation intensities with a 
cutoff below 5$\times\sigma_{PI}$. Contours start at 5~mK~$T_{\rm B}$ and run 
in steps of 10~mK~$T_{\rm B}$. The gamma-ray pulsar discovered by {\it Fermi} 
is marked as a star. The eastern shell, central branch and southern shell are 
sketched by blue dashed lines.}
\label{cta16cm}
\end{figure}

We show the new $\lambda$6~cm total intensity and polarisation maps in 
Fig.~\ref{cta16cm}. The polarisation intensity ($PI$) was obtained as 
$PI=\sqrt{U^2+Q^2-1.2\sigma_{U,\,Q}^2}$ following \citet{wk74} to correct for 
the positive noise offset. The polarisation angle ($\psi$) was calculated as 
$\psi=\frac{1}{2}{\rm atan}\frac{U}{Q}$. Here $\sigma_U$ and $\sigma_Q$ are the 
rms-noise of $U$ and $Q$. The rms-noise is 0.5~mK~$T_{\rm B}$ for $I$, and 
0.3~mK~$T_{\rm B}$ for $U$, $Q$, and $PI$, which are quite close to the 
theoretical expectations. 

Similar to previous radio observations of CTA~1 \citep{ssm81,plsr97}, a very 
bright semicircular shell roughly centred at 
${\rm RA}\approx0^{\rm h}6^{\rm m}$, and ${\rm Dec}\approx72\degr48\arcmin$ 
with a radius of about $50\arcmin$ clearly outlines half of the SNR. The 
opposite half of CTA~1 mainly manifests weak and diffuse emission, which 
characterizes the ``breakout" region. A bright arc protruding from the southern 
shell is clearly seen, which is called central branch hereafter. This 
corresponds to the ``bridge" named by \citet{plsr97}. The weak 
hooklike feature (just above the gamma-ray pulsar indicated in 
Fig.~\ref{cta16cm}) as reported by \citet{plsr97} connecting the top parts of 
the central branch and eastern shell can be identified. The strong compact source 
coinciding with the shell is planetary nebula NGC 40 
(${\rm RA}=0^{\rm h}13^{\rm m}2^{\rm s}$, 
${\rm Dec}=72\degr30\arcmin31\arcsec$), discussed in Appendix~\ref{sect_ngc40}. 

Strong polarised emission corresponding to the shell and the central branch was 
detected. The B-vectors follow the shell and the central branch. 
As shown below in Sect.~\ref{rmdet}, the intrinsic RM of CTA~1 is up to 
 40~rad~m$^{-2}$. This means the polarization angles at $\lambda$6~cm deviate 
less than $9\degr$ from their intrinsic values. Thus the B-vectors at 
$\lambda$6~cm shown in Fig.~\ref{cta16cm} fairly represent the orientation of 
the magnetic field.
Near NGC 40, the 
polarisation is much weaker than in the neighbouring area, which splits the 
shell into two parts (the eastern shell and the southern shell hereafter). This 
low-polarisation region is probably caused by a dense cloud pre-dating the 
supernova explosion, which drastically decreases the particle acceleration 
efficiency \citep{plsr97}. There is polarised emission connecting the central branch 
and the eastern shell. The average polarisation percentage ($PC$) at 
$\lambda$6~cm is about 37\% for the eastern shell, about 30\% for the central branch, 
and about 28\% for the southern shell.

\subsection{Effelsberg $\lambda$11~cm observations}
CTA~1 was observed by \citet{shs79} using the Effelsberg 100-m telescope at 
2695~MHz. The system temperature was about 100~K at that time. A new 
$\lambda$11~cm receiver was installed in 2005, which has a much lower system 
temperature of 17~K. We conducted observations of CTA~1 at $\lambda$11~cm using 
the new system in November and December 2005 and January 2006. Four 
coverages were obtained. The relevant observation parameters are listed 
in Table~\ref{obs}. 

The data were processed following standard procedures for Effelsberg continuum 
observations, which were the same used to reduce the Urumqi $\lambda$6~cm 
observations.

\begin{figure}[!htbp]
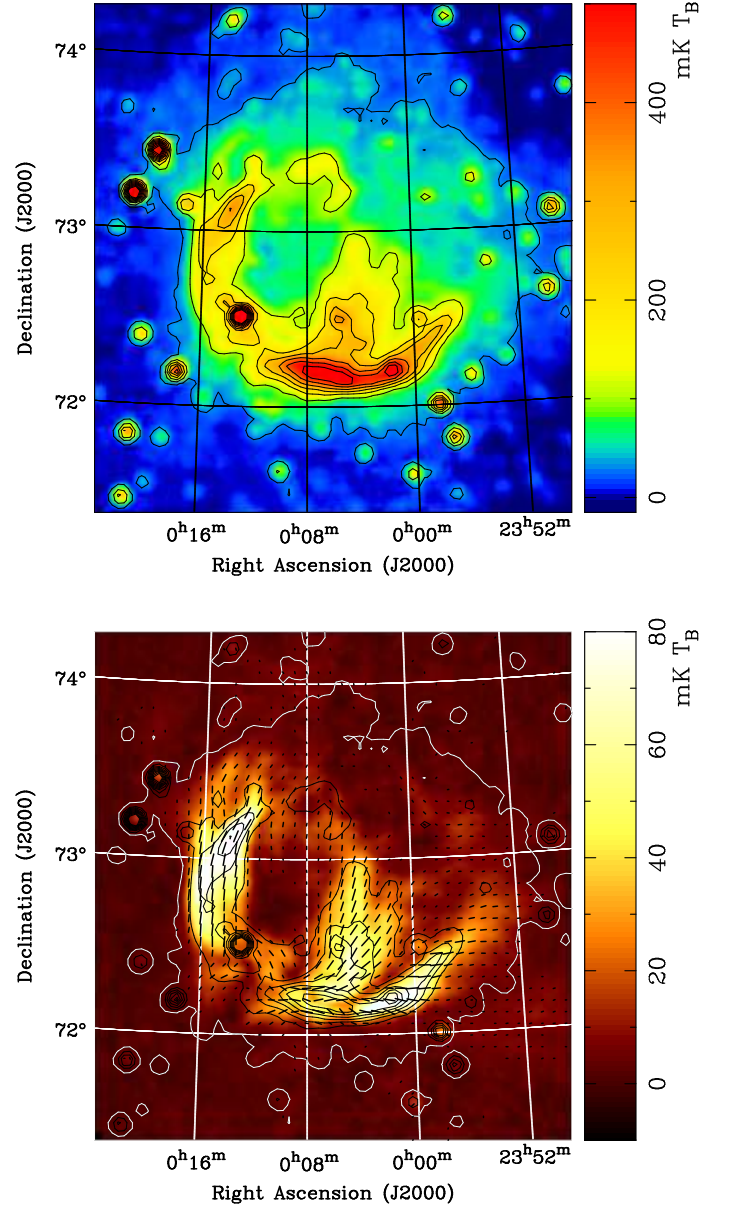

\centering
\resizebox{0.49\textwidth}{!}{\includegraphics[angle=-90]{cta1.11.i.ps}}\\[5mm]
\resizebox{0.49\textwidth}{!}{\includegraphics[angle=-90]{cta1.11.ipib.ps}}
\caption{Same as Fig.~\ref{cta16cm} but for Effelsberg $\lambda$11~cm maps.
The starting level is 30~mK~$T_{\rm B}$ and the interval is 80~mK~$T_{\rm B}$ 
for the contours.}
\label{cta111cm}
\end{figure}

The total intensity and polarisation maps of CTA~1 at $\lambda$11~cm are shown 
in Fig.~\ref{cta111cm}. Both the total intensity and the polarisation maps 
resemble the corresponding $\lambda$6~cm maps, but have a higher angular 
resolution. The breakout region can be tracked further to the north than in the 
$\lambda$6~cm map, indicating a very steep spectrum. 

A number of details can be revealed in the polarisation image. The northern end 
of the eastern shell and the central branch seem to split CTA~1 into two branches. 
Towards the low-polarisation region near NGC~40, we note several discrete small 
patches as the one at ${\rm RA}=0^{\rm h}11^{\rm m}$, 
${\rm Dec}=72\degr24\arcmin$, which is barely visible in the $\lambda$6~cm 
polarisation image. These could be fragments of a dense cloud disrupted after 
the passage of the shock wave from the supernova explosion as discussed by 
\citet{plsr97}. 

The average polarisation percentage is about 30\% for the 
eastern shell, 22\% for the central branch, and 26\% for the southern shell. 

\subsection{408~MHz and 1420~MHz total intensity data}
408~MHz and 1420~MHz maps were published by \citet{plsr97}, which were observed 
with the DRAO synthesis array, where the missing large-scale components were 
included from the 408~MHz all-sky survey \citep{hssw82} and 1420~MHz 
observations made with the Effelsberg telescope including data from the 
1420~MHz Stockert survey \citep{rei82} for the broadest structures. The maps 
have angular resolutions of $3\farcm5$ at 408~MHz and $1\arcmin$ at 1420~MHz. 
We use these data to construct a spectral index map together with the 
$\lambda$6~cm and $\lambda$11~cm observations.

\subsection{Effelsberg 1400~MHz polarisation data}
We extracted 1400~MHz polarisation data for the area of CTA~1 from an 
unpublished section of the Effelsberg Medium Latitude Survey (EMLS), which was 
described by \citet{ufr+98} and \citet{rei04}. The resolution is $9\farcm35$. 
The 1420~MHz total intensity map from the combined DRAO and Effelsberg 
observations is of high quality \citep{plsr97} and therefore not replaced by 
total intensity data from the EMLS. 

\begin{figure}[!htbp]
\centering
\resizebox{0.49\textwidth}{!}{\includegraphics[angle=-90]{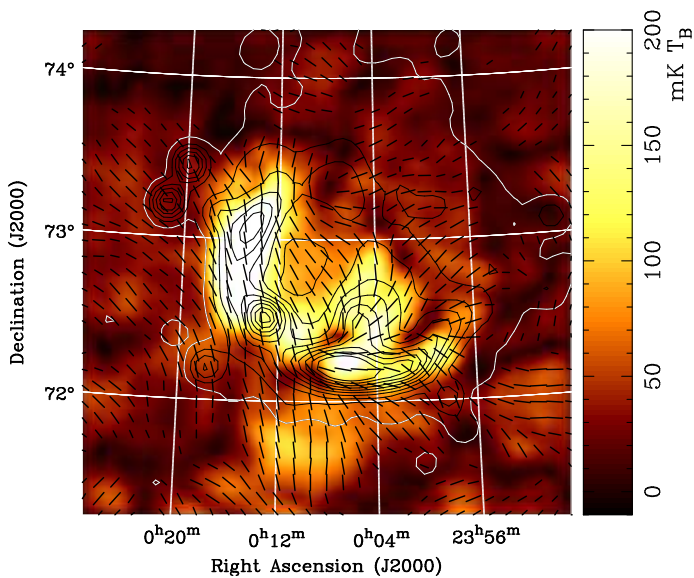}}
\caption{Same as Fig.~\ref{cta16cm} ({\it lower panel}) but for the EMLS 
$\lambda$21~cm map. Total intensity contour levels start at 
200~mK~$T_{\rm B}$ and run in steps of 200~mK~$T_{\rm B}$.}
\label{cta121cm}
\end{figure}

The $\lambda$21~cm polarisation map from the EMLS is shown in 
Fig.~\ref{cta121cm}. The polarisation distribution along the eastern shell is 
quite similar to that observed at $\lambda$6~cm and $\lambda$11~cm. For the 
southern shell, eastwards of ${\rm RA}\approx0^{\rm h}4^{\rm m}$, however, 
there is almost complete depolarisation. Parts of the central branch also experience 
large depolarisation. The average polarisation percentage is about 26\% for the 
eastern shell, 14\% for the central branch, and 10\% for the southern shell. 
Polarisation patches originating within the interstellar medium surround  
CTA~1 and most likely also exist along the line-of-sight to the SNR. They cause 
an overestimate of the $\lambda$21~cm polarisation percentage of CTA~1.

\section{Spectrum and rotation measure of CTA~1}

\subsection{Integrated flux densities and spectrum of CTA~1}
We used flux-density measurements of CTA~1 at 408~MHz, 1420~MHz, 2639~MHz, and 4800~MHz to 
constrain the spectrum of this SNR. Data at lower frequencies compiled by 
\citet{ssm81} were not used, because their very low angular resolution makes it 
difficult to assess the contribution of compact sources. 

Compact, point like sources visible in the total intensity maps at 408~MHz, 
1420~MHz, and 2639~MHz, were fitted by a 2D-elliptical Gaussian to subtract 
them. Most of the extragalactic sources are very weak at $\lambda$6~cm 
and confused with CTA~1. Therefore their flux densities cannot be measured from 
the $\lambda$6~cm map, and extrapolation from low frequency observations is the 
only way to estimate their contribution.
We obtained a map of sources at 1420~MHz and scaled it to 4.8~GHz 
using a spectral index of $\alpha=-0.9$ \citep[e.g.][]{zrrw03}. The spectral 
index is defined as $S_\nu\propto\nu^\alpha$ with $S_\nu$ being the flux 
density at frequency $\nu$. The scaled map was then smoothed to 
$9\farcm5$ and subtracted from the $\lambda$6~cm map. NGC~40 has been 
treated separately according to its spectrum shown in Fig.~\ref{ngc40}. 
A hyper-plane calculated from average values from the four map-corners was 
then subtracted from each map to set the surroundings of the SNR to zero. 

\begin{figure}[!htbp]
\centering
\resizebox{0.46\textwidth}{!}{\includegraphics[angle=-90]{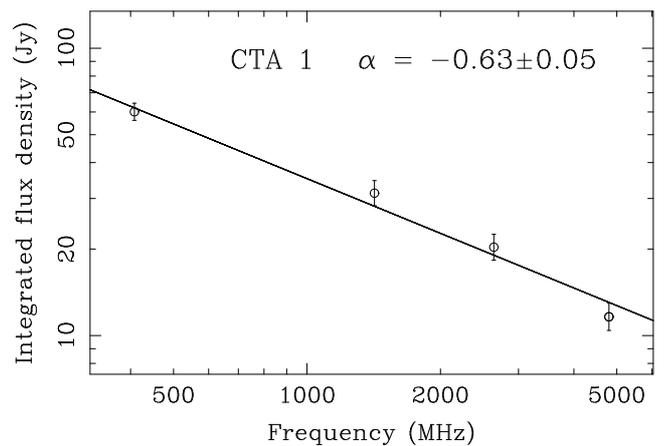}}
\caption{Fitted spectrum of integrated flux densities. }
\label{specS}
\end{figure}

We measured the integrated flux density of CTA 1 to be 11.6$\pm$1.2~Jy at 
$\lambda$6~cm and 20.3$\pm$2.0 at $\lambda$11~cm after discounting the 
contribution of extragalactic sources. We corrected the flux densities obtained 
by \citet{plsr97} accordingly and used integrated flux densities of 31$\pm$3~Jy 
and 60$\pm$4~Jy at 1420~MHz and 408~MHz, respectively. Fitting the flux density 
values at these four frequencies yields a spectral index of 
$\alpha=-0.63\pm0.05$ (Fig.~\ref{specS}), consistent with that reported by 
\citet{plsr97} of $\alpha=-0.57\pm0.006$.

As a crosscheck for the integrated flux density spectral index, we made 
TT-plots \citep{tpkp62} between $\lambda$6~cm and the other three frequencies 
(Fig.~\ref{ttp}). With this method the influence of a possible incorrect 
zero-level setting can be controlled. The TT-plot spectral index ($\beta$) 
from brightness temperatures relates to $\alpha$ as 
$\alpha=\beta+2$. As shown in Fig.~\ref{ttp}, the spectral index from 
TT-plots is $\beta=-2.58\pm0.03$ between $\lambda6$~cm and 408~MHz, 
$\beta=-2.61\pm0.03$ between $\lambda$6~cm and $\lambda$21~cm, and 
$\beta=-2.61\pm0.05$ between $\lambda$6~cm and $\lambda$11~cm. These values 
convincingly confirm the spectral index we obtained from the integrated flux 
densities. The group of outliers at 40--50~mK~$T_{\rm B}$ 
in the TT-plots in Fig.~\ref{ttp} are from the residuals when subtracting 
NGC~40 from the $\lambda$6~cm map.

\subsection{Spectrum steepening}

A steepening of the spectrum from the shell towards the breakout region was 
already reported by \citet{plsr97} based on the 408~MHz and 1420~MHz 
observations. The new $\lambda$6~cm and $\lambda$11~cm measurements allow us to 
study the spatial variations of the spectral index more accurately, because of 
the wider frequency range.

\begin{figure*}[!htbp]
\centering
\resizebox{0.32\textwidth}{!}{\includegraphics[angle=-90]{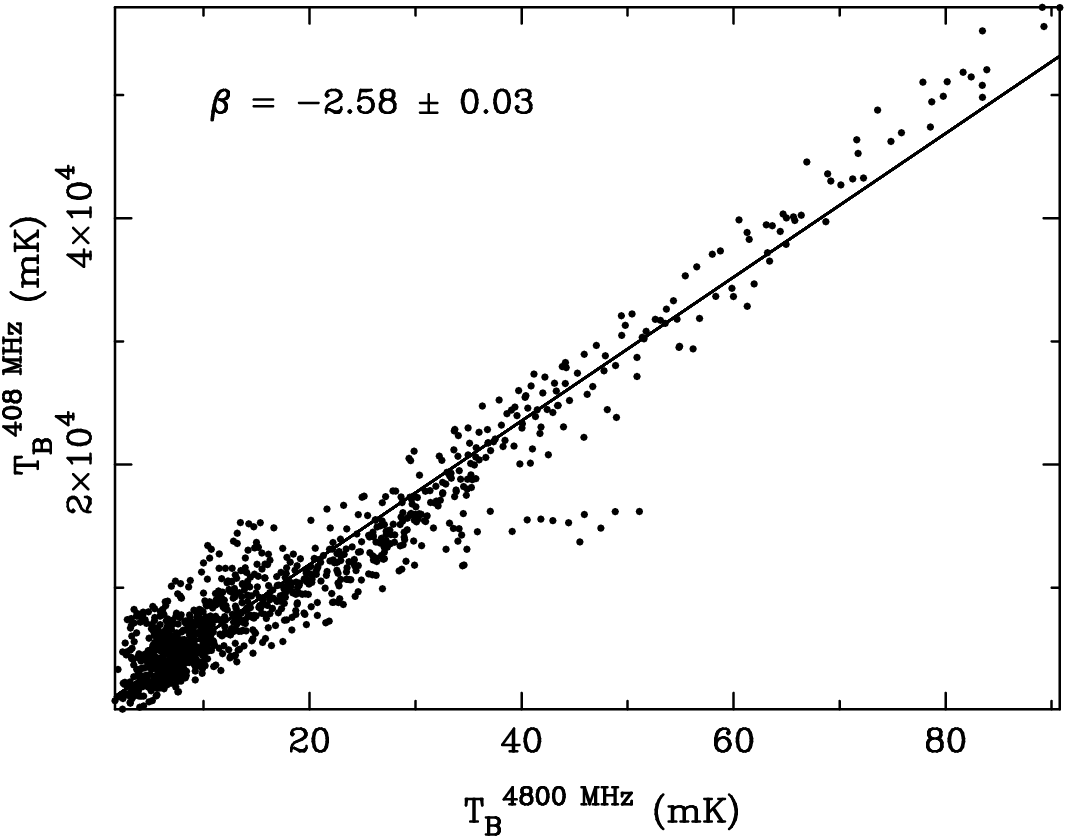}}
\resizebox{0.32\textwidth}{!}{\includegraphics[angle=-90]{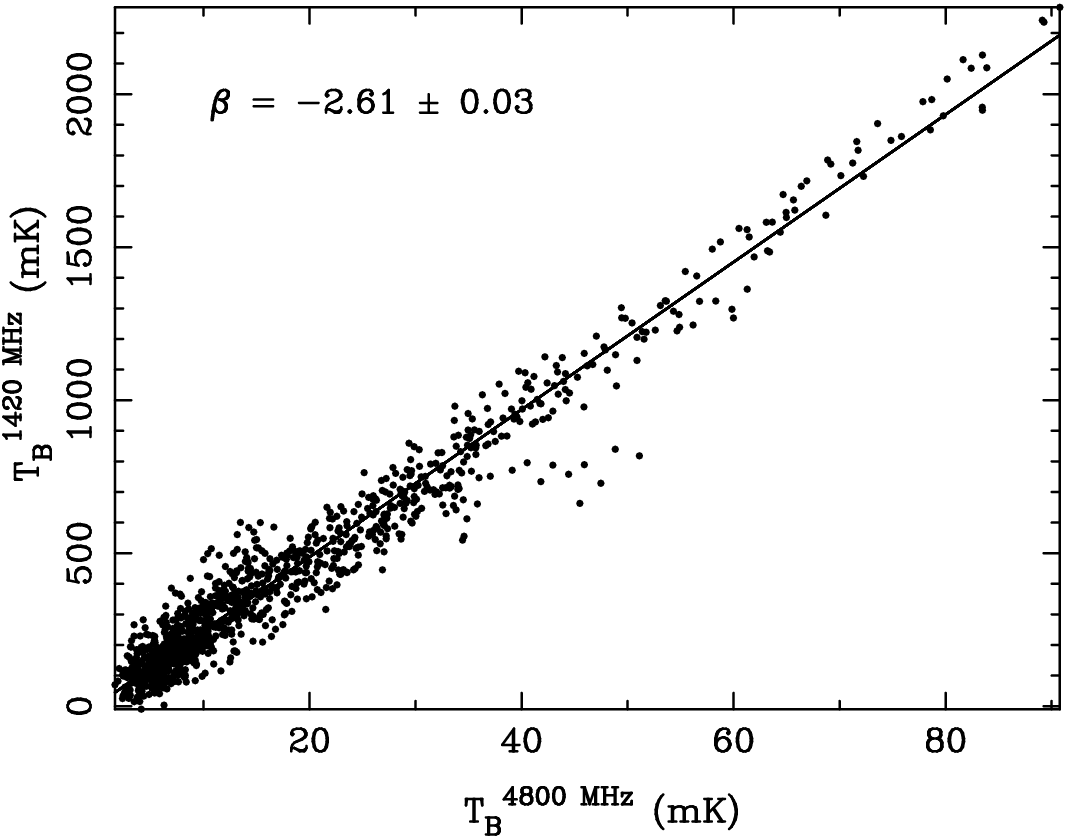}}
\resizebox{0.32\textwidth}{!}{\includegraphics[angle=-90]{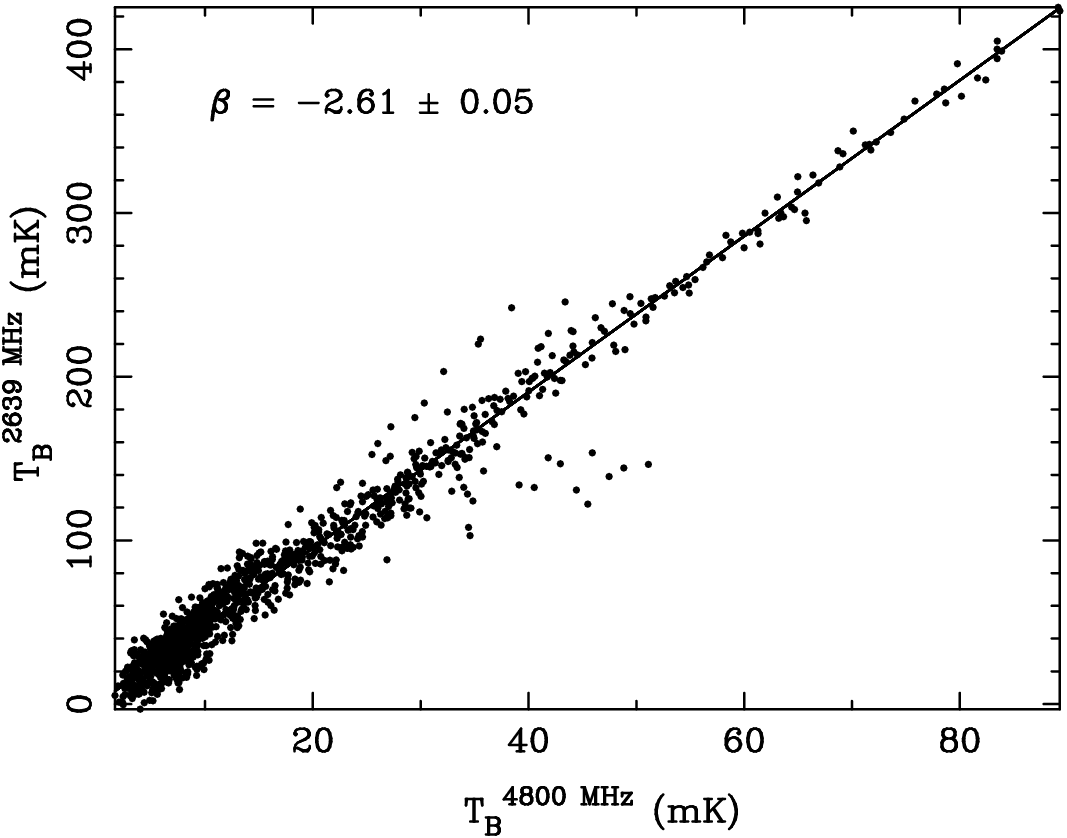}}
\caption{TT-plots between $\lambda$6~cm (4800~MHz) and other frequencies as indicated in the panels.}
\label{ttp}
\end{figure*}

\begin{figure*}[!htbp]
\centering
\resizebox{0.4\textwidth}{!}{\includegraphics[angle=-90]{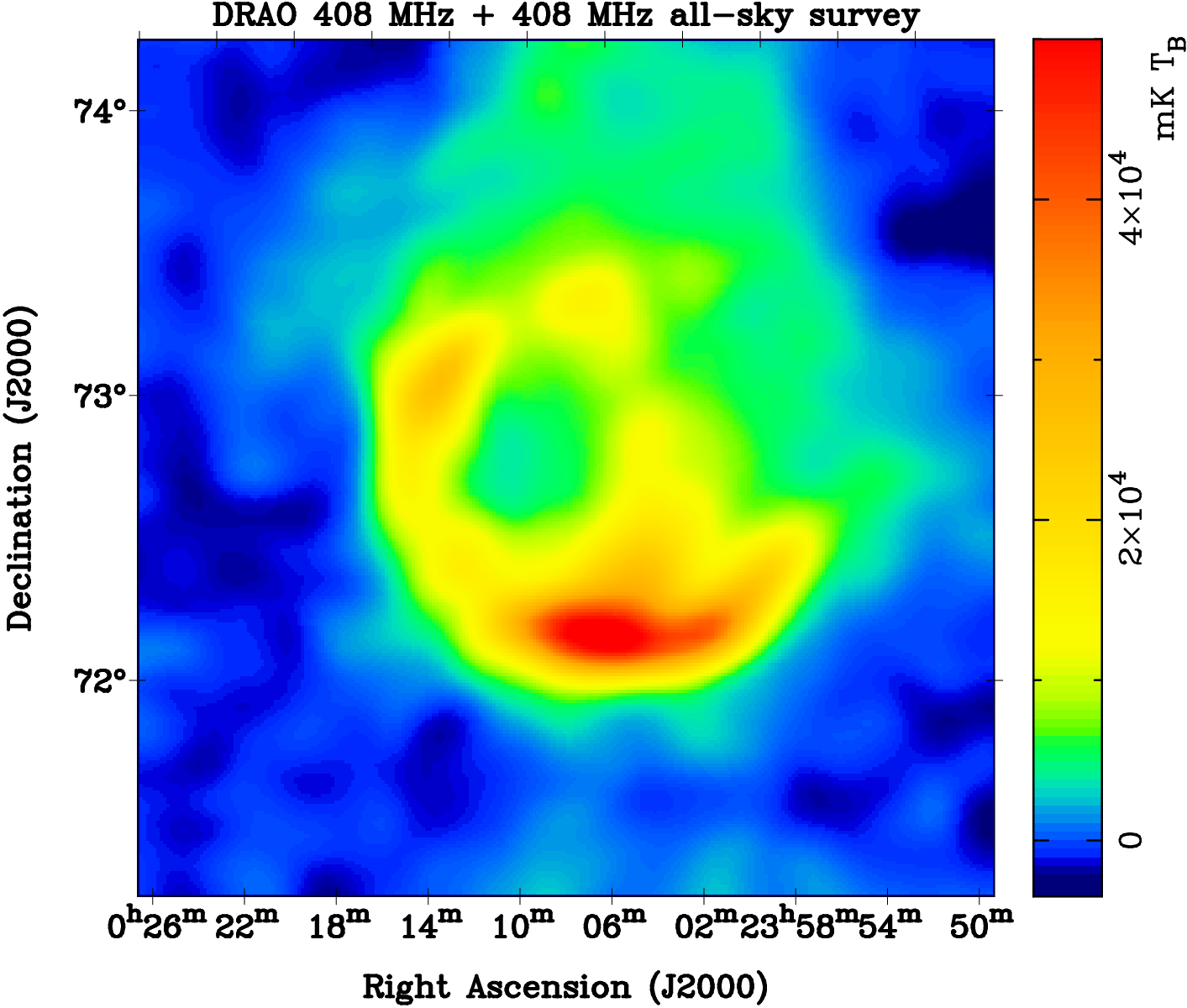}}
\resizebox{0.4\textwidth}{!}{\includegraphics[angle=-90]{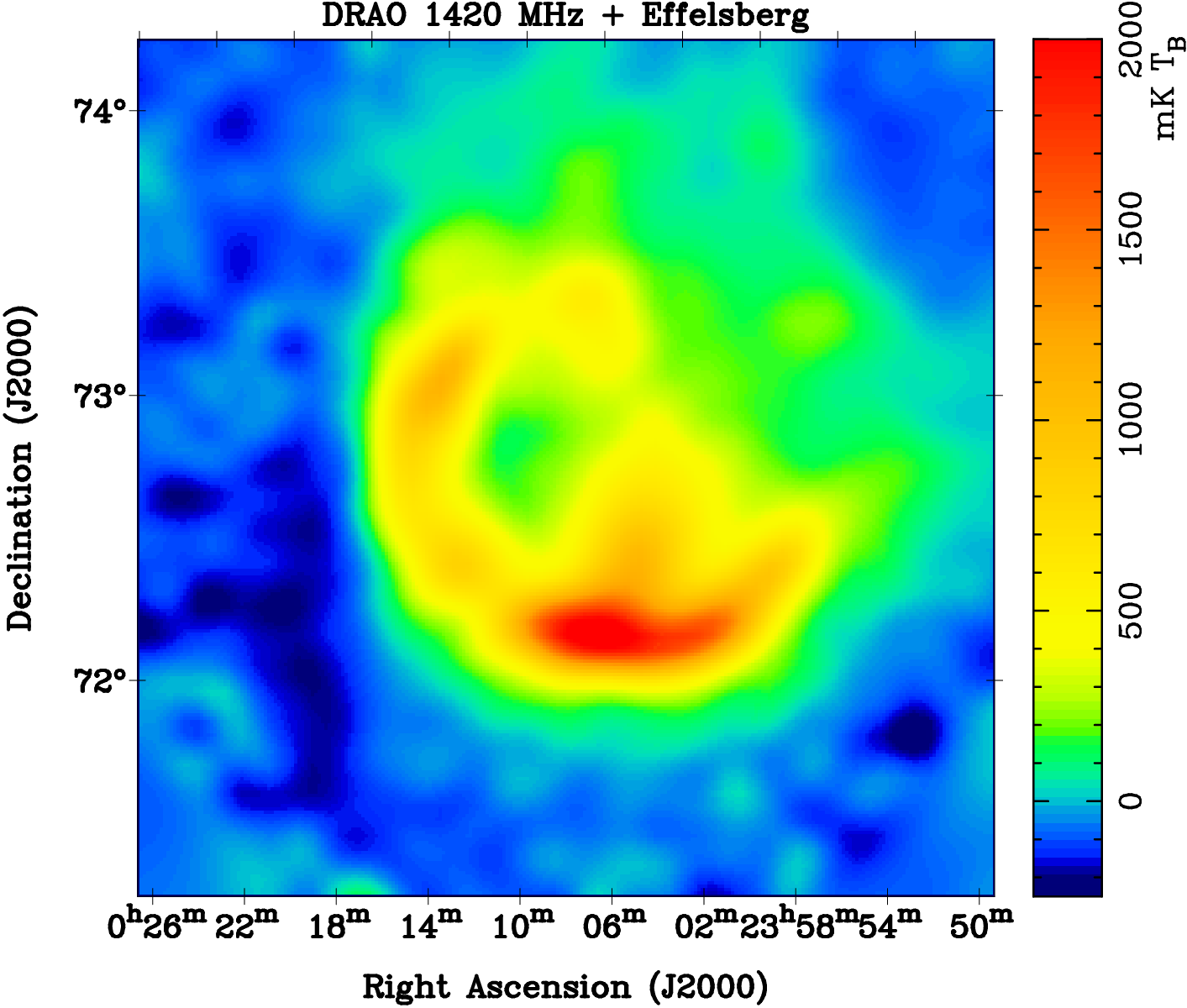}}
\\[5mm]
\resizebox{0.4\textwidth}{!}{\includegraphics[angle=-90]{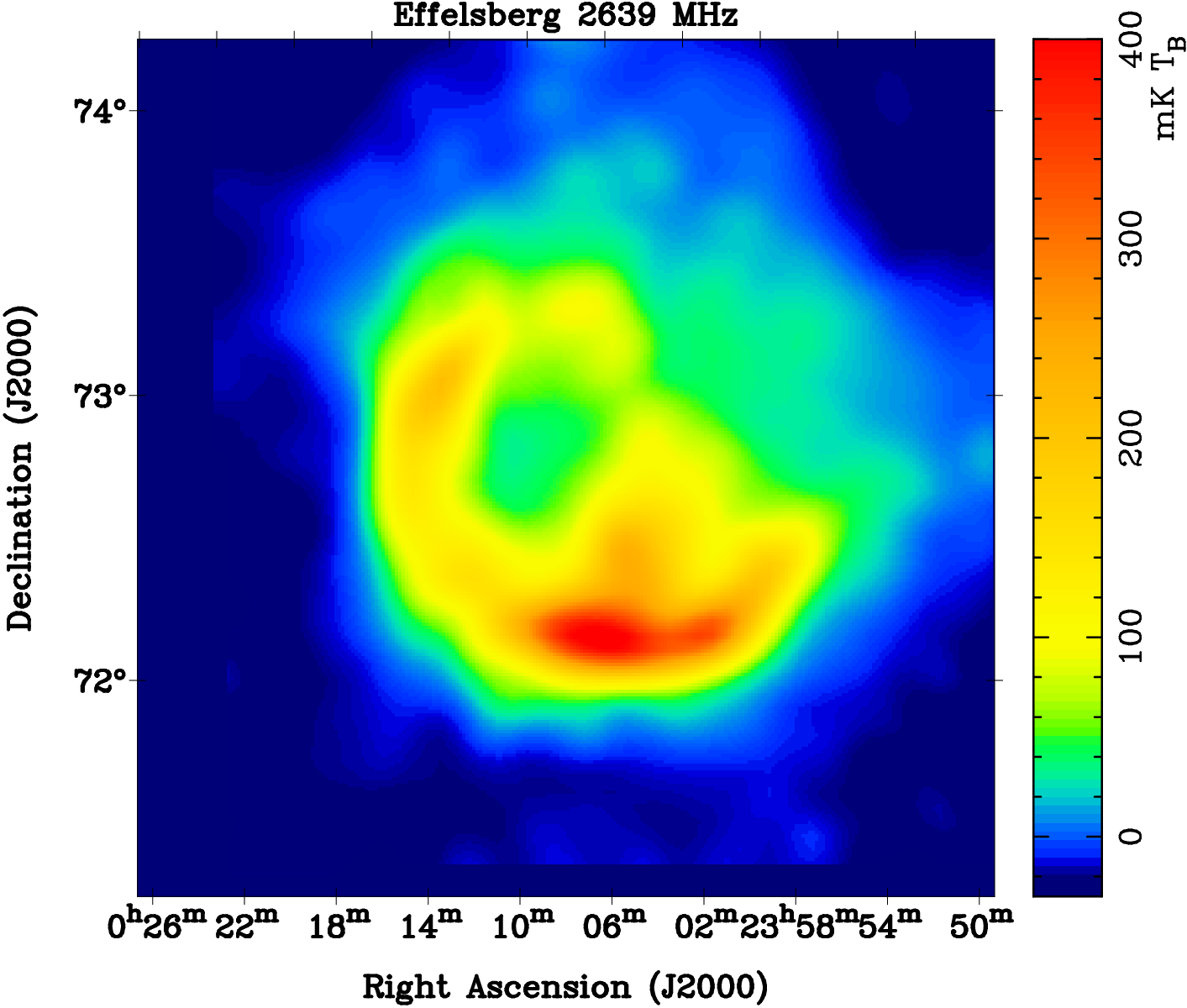}}
\resizebox{0.4\textwidth}{!}{\includegraphics[angle=-90]{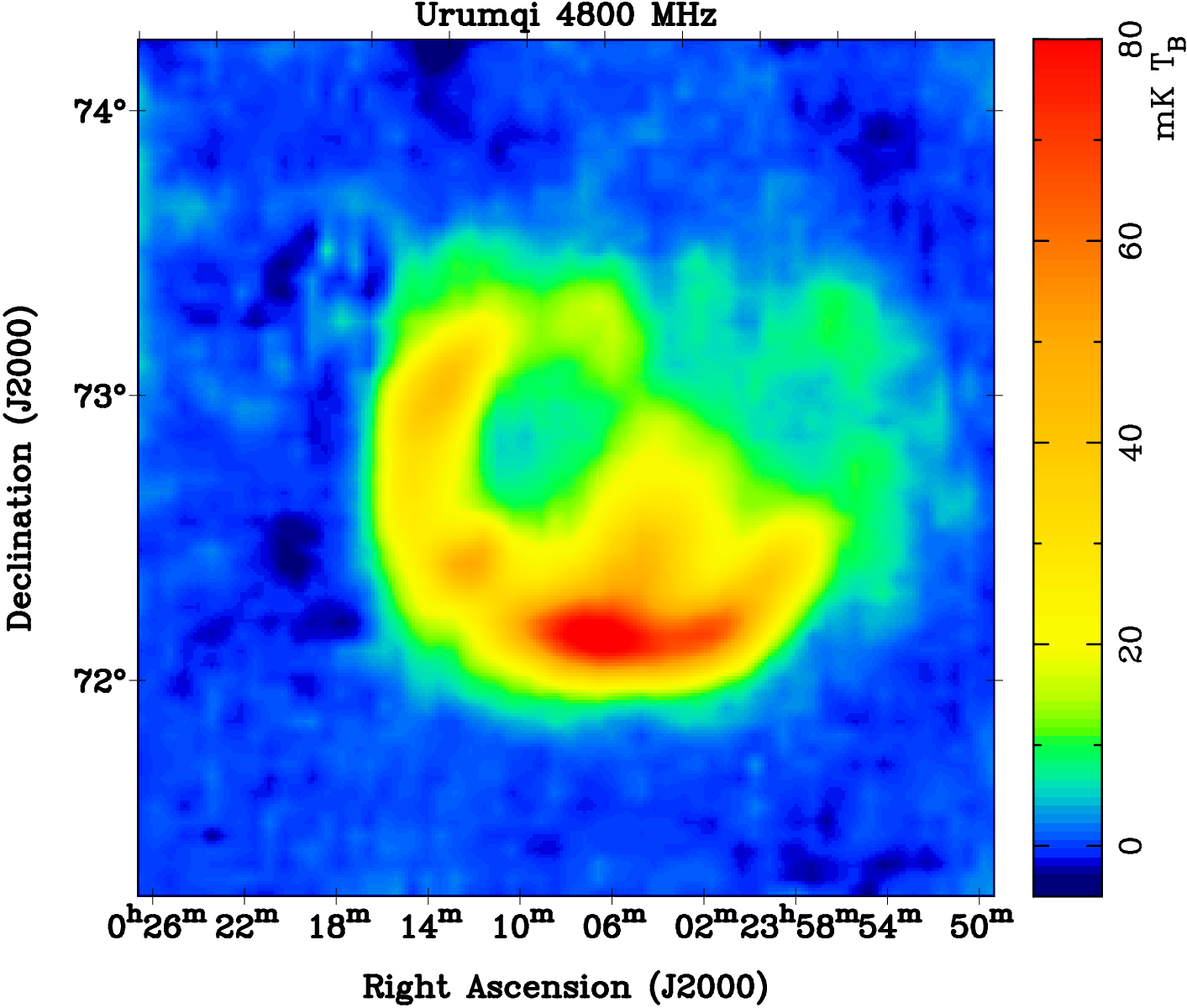}}
\caption{Total intensity maps at four frequencies. The sources have been 
subtracted and the zero-levels have been corrected. All the maps have a 
common resolution of $9\farcm5$. }
\label{i4f}
\end{figure*}

Towards the SNR, the observed intensity is a sum of the intrinsic emission from 
the SNR and the fluctuations of the diffuse interstellar medium on degree 
scales. To set up a common zero-level for all the maps, an offset has to be 
applied for each frequency before calculating a spectral index map of CTA~1. 
The source-removed maps at 408~MHz, 1420~MHz and 2639~MHz were smoothed to an 
angular resolution of $9\farcm5$. We assume the offset to be zero at 4.8~GHz. 
This is reasonable because the 
large-scale emission is very weak at the high-latitude of CTA~1. We then added 
or subtracted a constant offset from the maps at each of the other three 
frequencies to assure the TT-plots versus $\lambda$6~cm intersect at zero 
brightness temperature (Fig.~\ref{ttp}). Total intensities of 830~mK, 87~mK 
and 24~mK were subtracted from maps at 408~MHz, 1420~MHz and 2639~MHz, 
respectively, to properly settle the zero-levels. The adjusted maps are 
displayed in Fig.~\ref{i4f}. 

\begin{figure}
\centering
\resizebox{0.5\textwidth}{!}{\includegraphics[angle=-90]{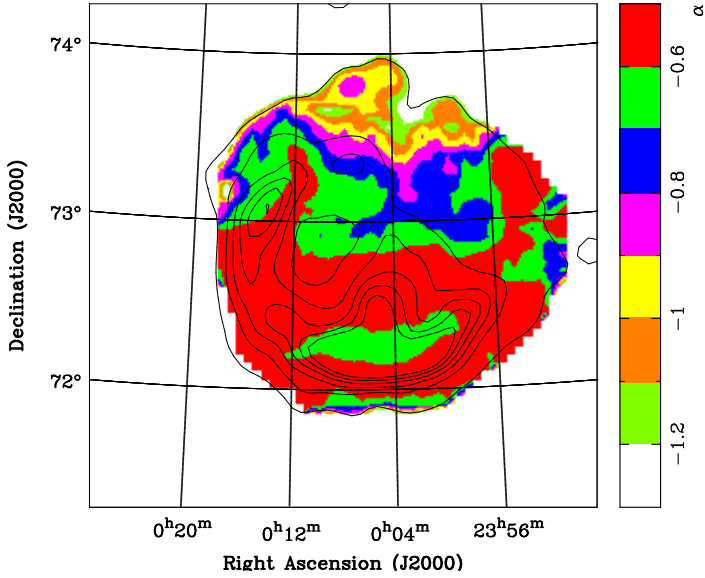}}
\caption{Spectral index map of CTA~1 calculated from the four maps between 
408~MHz and 4.8~GHz. The contours indicate the smoothed $\lambda$11~cm total 
intensities shown in Fig.~\ref{i4f}.}
\label{specM}
\end{figure}

For each pixel of the region encompassing CTA~1 from zero-level corrected maps 
at the four frequencies (Fig.~\ref{i4f}), we made a linear fit for 
intensities versus frequencies in logarithmic scale to obtain the spectral 
index map. For regions at ${\rm Dec}>73\degr30\arcmin$, only intensities at 
408~MHz, 1420~MHz and 2639~MHz were used since no diffuse emission from the 
SNR could be detected at 4.8~GHz exceeding the noise level. The result is shown 
in Fig.~\ref{specM}. The errors are generally less than 0.1 and become larger 
towards the very northern part of CTA~1, where the spectral index decreases to 
about $\alpha=-1$.   

Spectral index values are between $\alpha=-0.5$ and $\alpha=-0.65$ towards the 
shell and the central branch, and gradually become smaller towards the breakout region. 
The variation is up to about $\Delta\alpha=0.3$, which confirms the results by 
\citet{plsr97}.  

The mechanism for the spectrum steepening is not yet clear. 
It is widely accepted that cosmic-ray electrons experience diffusive shock 
acceleration (DSA) in SNR shock fronts and become relativistic. The spectral 
index of the synchrotron emission observed from these electrons can be written 
as $\alpha=-3/2(r-1)$, where $r$ is the shock compression ratio. To produce the 
observed spectral index, the compression ratio is about 3.5--4 for the shell, 
and 2.7--2.9 for the breakout region. A decrease of the compression ratio 
requires a decreasing shock Mach number, which is possible in case of a low gas 
density in the breakout region \citep{plsr97}. However, whether DSA is still 
important in the breakout region remains very uncertain. 

It could be the case that the steep-spectrum emission of the breakout region 
originates from higher-energy electrons than in the SNR shell because of a 
weaker magnetic field there. These high-energy electrons may have a steeper 
spectrum than the low-energy electrons due to synchrotron aging. They may have 
been accelerated and diffused out from the SNR shell. In this case a 
steepening at higher frequencies should be seen in the SNR spectra. However, no 
indication of a spectral break is found in the frequency range between 408~MHz 
and 4800~MHz, which means that any spectral break due to synchrotron aging 
should be at higher frequencies. Sensitive observations of CTA~1 at even higher 
frequencies are needed including measurements of the extended diffuse emission 
from the breakout region, although it is difficult to map such a large object 
with arcmin angular resolution.

\subsection{RM map of CTA~1}

\subsubsection{RM determination}\label{rmdet}

\begin{figure}[!htbp]
\resizebox{0.48\textwidth}{!}{\includegraphics[angle=-90]{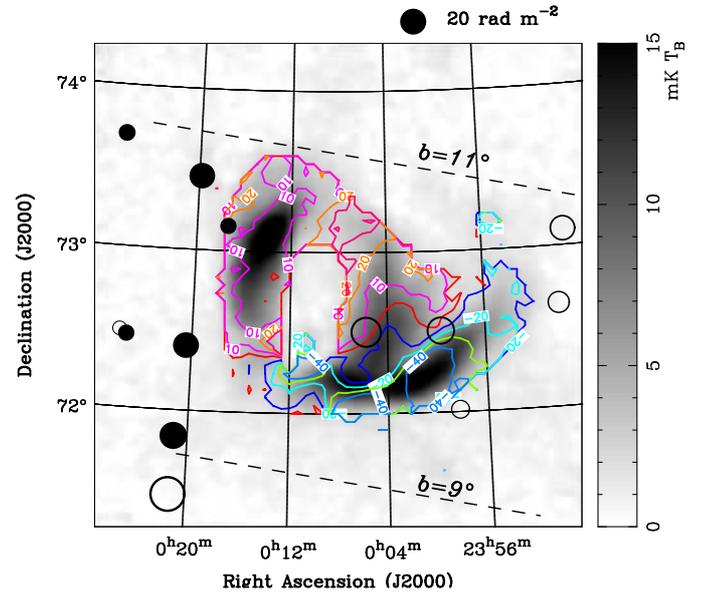}}
\caption{RM contours overlaid on the $\lambda$6~cm polarisation image. 
The contour levels are 0 (red), 10 (magenta), 20 (orange), and 30~rad~m$^{-2}$ 
(red+magenta) for 
positive RMs and $-$10 (blue), $-$20 (cyan), $-$30 (green+yellow), and $-$40~rad~m$^{-2}$ (blue+cyan) for 
negative RMs. 
The filled (open) circles represent positive (negative) RMs of extragalactic 
sources taken from \citet{tss09}. The dashed line marks Galactic latitude 
$b=9\degr$ and $b=11\degr$.}
\label{rmmap}
\end{figure}

We smoothed the $\lambda$11~cm $U$ and $Q$ data to an angular resolution of 
$9\farcm5$ and derived a map of polarisation angles. According to the 
polarisation angles at $\lambda$6~cm and $\lambda$11~cm, we calculated RMs 
as, 
\begin{equation}
{\rm RM}=\frac{\psi_{\rm 6\,cm}-\psi_{\rm 11\,cm}}
              {\lambda^2_{\rm 6\,cm}-\lambda^2_{\rm 11\,cm}}
\end{equation}
RM is proportional to the integral of the thermal electron density multiplied 
by the magnetic field parallel to the line-of-sight.
Pixels with brightness temperatures less than $5\times\sigma_{PI}$ were not 
included. The results are shown in Fig.~\ref{rmmap}. RMs of extragalactic 
sources in the CTA~1 area were taken from the catalogue by \citet{tss09} 
and displayed in Fig~\ref{rmmap}. Their RMs range between $\pm$40~rad~m$^{-2}$ 
and are much smaller than the RM ambiguity of 348~rad~m$^{-2}$ between 
$\lambda$6~cm and $\lambda$11~cm. Therefore we always use the minimum 
polarisation angle difference to calculate RM. 

The RM distribution of CTA~1 exhibits a clear pattern. Towards the eastern 
shell and the central branch, RMs are always positive, and always negative along the 
southern shell. The average of RMs for the eastern shell and the central branch is 
about +10~rad~m$^{-2}$ and +15~rad~m$^{-2}$. For the southern shell, 
the absolute values of RMs are as large as about 50~rad~m$^{-2}$. 
The RM map by \citet{ssm81} based on the Effelsberg $\lambda$11~cm and 
$\lambda$18~cm observations qualitatively agrees with our result.
Interestingly the RMs of extragalactic sources show a similar pattern as CTA~1, 
but extend to a much larger region as discussed in Sect.~\ref{fs}.
 
\subsubsection{A Faraday screen in front of CTA~1}\label{fs}

\begin{figure}[!htbp]
\resizebox{0.45\textwidth}{!}{\includegraphics[angle=-90]{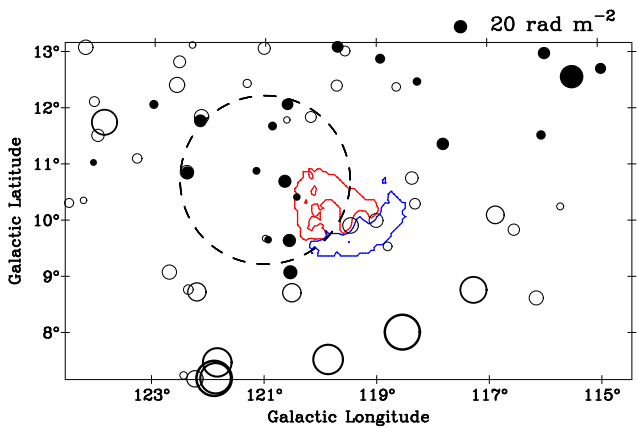}}
\caption{Similar to Fig.~\ref{rmmap} but for a larger field in Galactic 
coordinates. The contours outline positive (+10~rad~m$^{-2}$, red) and negative 
(-10~rad~m$^{-2}$, blue) RMs. The circle marks the approximate boundary of the 
Faraday screen.}
\label{egsrm}
\end{figure}

The RM pattern of CTA~1 as well as that of extragalactic sources 
(Fig.~\ref{rmmap}) strongly suggests, that there is a Faraday screen in the 
direction of the eastern shell and the central branch of CTA~1. The Faraday screen 
exceeds the size of CTA~1 as it can be clearly seen from a larger field 
containing RMs from extragalactic sources (Fig.~\ref{egsrm}). Therefore the 
Faraday screen should be located in front of CTA~1, which has a distance of 
about 1.4~kpc. Its centre is roughly at $l=121\degr$, $b=11\degr$ with a poorly 
constrained size of about $3\degr$. To estimate the RM of the Faraday screen, 
we first need to investigate RMs from the SNR and the Galactic foreground.

\begin{figure}[!htbp]
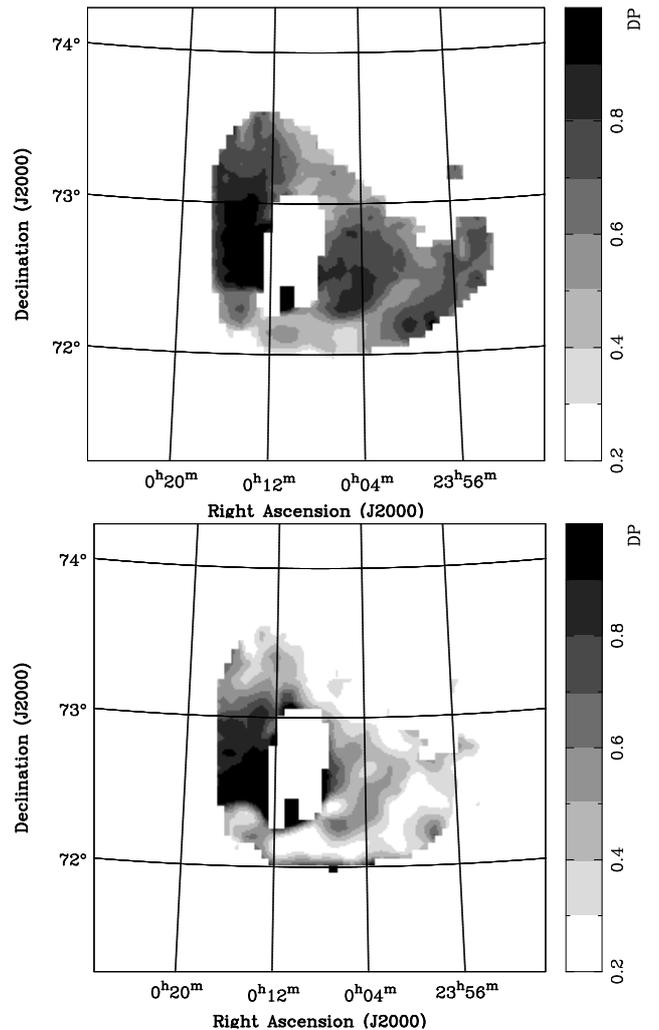

\resizebox{0.45\textwidth}{!}{\includegraphics[angle=-90]{cta1.dp.11.ps}}
\resizebox{0.45\textwidth}{!}{\includegraphics[angle=-90]{cta1.dp.21.ps}}
\caption{Relative depolarisation relative to $\lambda$6~cm at $\lambda$11~cm 
({\it upper panel}) and $\lambda$21~cm ({\it lower panel}).}
\label{dpmap}
\end{figure}

The intrinsic RMs of CTA~1 can be estimated from depolarisation. Polarised 
emission originating from different depth experiences different Faraday 
rotation, and summing up all the emission components along the line-of-sight 
may partly cancel polarised emission components. We define the relative 
depolarisation ($DP$) at $\lambda$11~cm and $\lambda$21~cm as 
$DP=PC/PC_{\lambda6~{\rm cm}}$, where $PC$ is the polarisation percentage. The $\lambda$11~cm and $\lambda$21~cm 
depolarisation maps are shown in Fig.~\ref{dpmap}. Following \citet{sbs+98} 
wavelength-dependent depolarisation is related to RM as,
\begin{equation}
\displaystyle{DP_\lambda=\frac{\sin(2|{\rm RM}|\lambda^2)}
{\sin(2|{\rm RM}|\lambda_0^2)}\times\frac{\lambda_0^2}{\lambda^2}},
\end{equation}
where we take $\lambda_0=6.25$~cm (4.8~GHz). 

The average depolarisation is about 0.9, 0.8, and 0.8 for the eastern shell, 
the central branch, and the southern shell, respectively. This implies a small amount 
of depolarisation at $\lambda$11~cm. At $\lambda$21~cm, the average 
depolarisation is about 0.9 for the eastern shell, which means an intrinsic 
absolute RM of 10~rad~m$^{-2}$. The average depolarisation of about 0.6 towards 
the central branch requires an absolute RM value of of about 18~rad~m$^{-2}$. Towards 
the southern shell except for several fragments, there is complete 
depolarisation at $\lambda$21~cm. The average depolarisation of the fragments 
is about 0.3, implying an absolute RM value larger than about 25~rad~m$^{-2}$. 

The intrinsic RMs of CTA~1 are negative following the regular Galactic 
magnetic field direction in this area. The observed positive RMs towards the 
eastern shell and the central branch are caused by the Faraday screen in front 
of CTA~1.

Based on the 3D-emission models of the Milky Way by \citet{srwe08} and 
\citet{sr10}, a RM from the diffuse interstellar medium up to the 1.4~kpc 
distance of CTA~1 was calculated to be about  $-17\pm8$~rad~m$^{-2}$. 
Unfortunately the newly discovered gamma-ray pulsar by \textit{Fermi} 
\citep{aaa+08} does not yet have a radio detection and thus a measured RM. 
However, the pulsar B0105+68, about $6\degr$ offset from the centre of CTA~1 
has a distance of about 2.6~kpc and a RM of $-46$~rad~m$^{-2}$ \citep{mwkj03}. 
A linear interpolation to 1.4~kpc distance yields a RM of $-24$~rad~m$^{-2}$, 
consistent with that from the 3D-emission models. 

The RM of the Faraday screen can be calculated from the observed RMs towards 
the eastern shell and the central branch of CTA~1 by subtracting the intrinsic CTA~1 
values and the Galactic foreground contribution. The result is about 
37--50~rad~m$^{-2}$. In the following we use a mean value of about 
45~rad~m$^{-2}$ for the Faraday screen. For the southern shell, which is not 
influenced by the Faraday screen, the intrinsic RMs plus that from the Galactic 
foreground roughly agree with the observed ones. 

The Faraday screen will cause a rotation of the polarisation angles at 
$\lambda$21~cm by more than $100\degr$. However, we cannot find a 
correspondence with the EMLS data. This implies that most of the polarised 
emission observed at $\lambda$21~cm originates from regions closer than the 
Faraday screen.  

Assuming the Faraday screen to be spherical with a mean size of about $3\degr$, 
the regular magnetic field parallel to the line-of-sight can be estimated as 
$B_\parallel=3.6\,{\rm RM}/\sqrt{I_{{\rm H}\alpha}d}$. Here $I_{{\rm H}\alpha}$
is the H$\alpha$ intensity in R, $d$ is the distance in pc, the electron 
temperature is assumed to be 8000~K, and the extinction is neglected. There is 
no enhanced H$\alpha$ emission visible towards the Faraday screen in the 
Wisconsin H-Alpha Mapper northern sky survey \citep{hrt+03}. The average and 
the fluctuations of the H$\alpha$ intensity are about 4~R and 1~R, 
respectively, which implies an upper limit of about 7~R for the Faraday screen. 
For an assumed distance of 500(1000)~pc, the depth of the screen along 
line-of-sight is about 26(52)~pc and the lower limit of the regular magnetic 
field parallel to the line-of-sight is about 2.7(1.9)~$\mu$G.

\subsubsection{RM structure functions}

\citet{sim92} found that CTA~1 is a site of enhanced interstellar plasma 
turbulence based on a structure function analysis for a number of RMs of 
extragalactic sources at line-of-sights through and outside the SNR. 
The RM map we obtained above allows us to compare the structure function of RMs 
from CTA~1 directly with that of RMs from extragalactic sources by 
\citet{sim92} and \citet{tss09}. 

\begin{figure}[!htbp]
\resizebox{0.45\textwidth}{!}{\includegraphics[angle=-90]{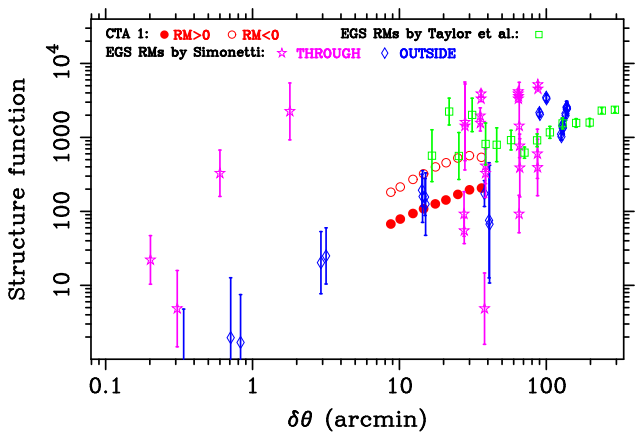}}
\caption{Structure functions for RMs of extragalactic sources taken from 
\citet{tss09}, the southern shell of CTA~1 (negative RMs), and the eastern 
shell and the brigde of CTA~1 (positive RMs). The squared RM differences for 
extragalactic sources at line-of-sights through and outside CTA~1 from 
\citet{sim92} are also shown.}
\label{sf}
\end{figure}

The second order structure function is a frequently used tool to infer 
turbulent properties of the interstellar medium \citep[e.g.][]{sr09,sts11}, 
which is defined as,
\begin{equation}
D(\delta\theta)=<[{\rm RM}(\theta)-{\rm RM}(\theta+\delta\theta)]^2>.
\end{equation}
Here $\delta\theta$ is the angular separation and $<\cdots>$ stands for the 
ensemble average. We calculated the structure functions from the RM map of 
CTA~1 for the southern shell, where RM is negative, and for the eastern shell 
and the central branch, where RM is positive (see Figs.~\ref{rmmap} and \ref{egsrm}).  
We also obtained the structure function for RMs of extragalactic sources 
shown in Fig.~\ref{egsrm}. Note that several sources coinciding with CTA~1 
(Fig.~\ref{egsrm}) were excluded in order to compare with the 
RM structure function for extragalactic sources with line-of-sights outside of 
CTA~1 by \citet{sim92}. The results are shown in Fig.~\ref{sf}. 

The structure functions for both positive and negative RMs of CTA~1 can be well 
fitted by a power law with a slope of about 0.85 in the range of 
$10\arcmin\leq\delta\theta\leq30\arcmin$, where $10\arcmin$ reflects the 
angular resolution and $30\arcmin$ is about the width of the shell. Note 
that the RMs of CTA~1 consist of contributions from the SNR and the Galactic 
foreground for the southern shell. For the eastern shell and the central branch, the 
contribution from the Faraday screen adds, which does not change the slope of 
the structure function. This indicates that the magnetic field in the Faraday 
screen along the line-of-sight is very regular, so that no additional 
fluctuations were introduced. The different amplitudes of the two structure 
functions stem from the different absolute values of RM, which are also 
reflected in the depolarisation properties (see Sect.~4.2).   

The squared RM differences of extragalactic sources at line-of-sights through 
and outside CTA~1 from \citet{sim92} are also shown in Fig.~\ref{sf}. At 
larger scales of $\delta\theta\gtrsim20\arcmin$, the amplitudes of RM structure 
functions for all extragalactic sources are similar. At smaller scales of 
$\delta\theta\lesssim3\arcmin$, the squared RM differences for extragalactic 
sources at line-of-sights through CTA~1 are larger than outside CTA~1. 
Therefore \citet{sim92} proposed that CTA~1 induces enhanced plasma turbulence 
with an outer scale in the range of $3\arcmin$--$20\arcmin$. As indicated in 
Fig.~\ref{sf}, the amplitude of the RM structure function for the shell and 
central branch regions of CTA~1 roughly agrees with that for extragalactic sources at 
line-of-sights outside CTA~1, when extrapolated to smaller scales. This 
implies that the scenario of enhanced plasma fluctuation proposed by 
\citet{sim92} is limited at scales below about $3\arcmin$. Clearly more data 
are needed to settle this issue.

\section{Summary}
We performed new $\lambda$6~cm and $\lambda$11~cm continuum observations of 
the shell-type SNR CTA~1. We obtained integrated flux densities of 
11.6$\pm$1.2~Jy at $\lambda$6~cm and 20.3$\pm$2.0~Jy at $\lambda$11~cm after 
subtraction of the contributions from compact extragalactic sources. Together 
with values at 408~MHz and 1420~MHz we derived a spectral index of 
$\alpha=-0.63\pm0.05$, which was confirmed by TT-plots. We also obtained a 
spectral index map based on data at 408~MHz, 1420~MHz, 2639~MHz, and 4800~MHz. 
The map shows a spectral index variation up to about $\Delta\alpha=0.3$ from 
the shell towards the breakout region. The reason for the spectrum steepening 
needs further investigation.  

We obtained a RM map using polarisation data at $\lambda$6~cm and 
$\lambda$11~cm. A clear pattern of negative RMs towards the southern part and 
positive ones towards the eastern part can be seen from the RM map. We argued 
that there exists a Faraday screen of roughly $3\degr$ in size located in front 
of CTA~1 covering its northeastern part. The RM of the Faraday screen is about 45~rad~m$^{-2}$. Assuming 
a distance of 500(1000)~pc for the Faraday screen the lower limit for the 
regular magnetic field parallel to the line-of-sights is 2.7(1.9)~$\mu$G.

We calculated RM structure functions for different parts of CTA~1 and could 
not find an influence by the foreground Faraday screen, which indicates a
very regular enhanced magnetic field.   

\begin{acknowledgements}
We thank the staff of the Urumqi Observatory of NAOC for qualified assistance 
with the observations. In particular we like to thank Otmar Lochner for
the construction of the $\lambda$6\ cm system and its installation and
Maozheng Chen and Jun Ma for their help during the installation of the
receiver and subsequent maintenance. We are very grateful to Dr. Peter
M\"uller for the software development needed to make mapping observations 
at the Urumqi telescope possible. The MPG and the NAOC supported the 
construction of the Urumqi $\lambda$6\ cm receiving system by special funds.
The Chinese survey team is supported by the National Natural Science foundation of
China (10773016, 10821061, and 10833003), the National Key Basic Research 
Science Foundation of China (2007CB815403), and the Partner group of the MPIfR 
at NAOC in the frame of the exchange program between MPG and CAS for many
bilateral visits. XHS likes to thank the MPG and Prof. Michael Kramer for financial 
support during his stay at the MPIfR. We are grateful to Prof. Ernst F\"urst for
help with the $\lambda$11\ cm observations and critical reading of the manuscript. Based on observations with the 100-m telescope of the MPIfR (Max-Planck-Institut f\"ur Radioastronomie) at Effelsberg. We thank Prof. Dr. Tom 
Landecker for providing us the 408~MHz and 1420~MHz data. We thank the referee 
Prof. John Dickel for his instructive and helpful comments. 
\end{acknowledgements}

\bibliographystyle{aa}
\bibliography{/homes/xhsun/bibtex}

\begin{thebibliography}{49}
\expandafter\ifx\csname natexlab\endcsname\relax\def\natexlab#1{#1}\fi

\bibitem[{{Abdo} {et~al.}(2008){Abdo}, {Ackermann}, {Atwood}, {Baldini},
  {Ballet}, {Barbiellini}, {Baring}, {Bastieri}, {Baughman}, {Bechtol},
  {Bellazzini}, {Berenji}, {Blandford}, {Bloom}, {Bogaert}, {Bonamente},
  {Borgland}, {Bregeon}, {Brez}, {Brigida}, {Bruel}, {Burnett}, {Caliandro},
  {Cameron}, {Caraveo}, {Carlson}, {Casandjian}, {Cecchi}, {Charles},
  {Chekhtman}, {Cheung}, {Chiang}, {Ciprini}, {Claus}, {Cohen-Tanugi},
  {Cominsky}, {Conrad}, {Cutini}, {Davis}, {Dermer}, {de Angelis}, {de Palma},
  {Digel}, {Dormody}, {do Couto e Silva}, {Drell}, {Dubois}, {Dumora},
  {Edmonds}, {Farnier}, {Focke}, {Fukazawa}, {Funk}, {Fusco}, {Gargano},
  {Gasparrini}, {Gehrels}, {Germani}, {Giebels}, {Giglietto}, {Giordano},
  {Glanzman}, {Godfrey}, {Grenier}, {Grondin}, {Grove}, {Guillemot}, {Guiriec},
  {Harding}, {Hartman}, {Hays}, {Hughes}, {J{\'o}hannesson}, {Johnson},
  {Johnson}, {Johnson}, {Johnson}, {Kamae}, {Kanai}, {Kanbach}, {Katagiri},
  {Kawai}, {Kerr}, {Kishishita}, {Kiziltan}, {Kn{\"o}dlseder}, {Kocian},
  {Komin}, {Kuehn}, {Kuss}, {Latronico}, {Lemoine-Goumard}, {Longo}, {Lonjou},
  {Loparco}, {Lott}, {Lovellette}, {Lubrano}, {Makeev}, {Marelli}, {Mazziotta},
  {McEnery}, {McGlynn}, {Meurer}, {Michelson}, {Mineo}, {Mitthumsiri},
  {Mizuno}, {Moiseev}, {Monte}, {Monzani}, {Morselli}, {Moskalenko}, {Murgia},
  {Nakamori}, {Nolan}, {Nuss}, {Ohno}, {Ohsugi}, {Okumura}, {Omodei},
  {Orlando}, {Ormes}, {Ozaki}, {Paneque}, {Panetta}, {Parent}, {Pelassa},
  {Pepe}, {Pesce-Rollins}, {Piano}, {Pieri}, {Piron}, {Porter}, {Rain{\`o}},
  {Rando}, {Ray}, {Razzano}, {Reimer}, {Reimer}, {Reposeur}, {Ritz},
  {Rochester}, {Rodriguez}, {Romani}, {Roth}, {Ryde}, {Sadrozinski}, {Sanchez},
  {Sander}, {Parkinson}, {Schalk}, {Sellerholm}, {Sgr{\`o}}, {Siskind},
  {Smith}, {Smith}, {Spandre}, {Spinelli}, {Starck}, {Strickman}, {Suson},
  {Tajima}, {Takahashi}, {Takahashi}, {Tanaka}, {Thayer}, {Thayer}, {Thompson},
  {Thorsett}, {Tibaldo}, {Torres}, {Tosti}, {Tramacere}, {Usher}, {Van Etten},
  {Vilchez}, {Vitale}, {Wang}, {Watters}, {Winer}, {Wood}, {Yasuda}, {Ylinen},
  \& {Ziegler}}]{aaa+08}
{Abdo}, A.~A., {Ackermann}, M., {Atwood}, W.~B., {et~al.} 2008, Science, 322,
  1218

\bibitem[{{Becker} {et~al.}(1991){Becker}, {White}, \& {Edwards}}]{bwe91}
{Becker}, R.~H., {White}, R.~L., \& {Edwards}, A.~L. 1991, \apjs, 75, 1

\bibitem[{{Condon} {et~al.}(1998){Condon}, {Cotton}, {Greisen}, {Yin},
  {Perley}, {Taylor}, \& {Broderick}}]{ccg+98}
{Condon}, J.~J., {Cotton}, W.~D., {Greisen}, E.~W., {et~al.} 1998, \aj, 115,
  1693

\bibitem[{{Emerson} \& {Gr\"ave}(1988)}]{eg88}
{Emerson}, D.~T., \& {Gr\"ave}, R. 1988, \aap, 190, 353

\bibitem[{{Fesen} {et~al.}(1981){Fesen}, {Blair}, {Kirshner}, {Gull}, \&
  {Parker}}]{fbk+81}
{Fesen}, R.~A., {Blair}, W.~P., {Kirshner}, R.~P., {Gull}, T.~R., \& {Parker},
  R.~A.~R. 1981, \apj, 247, 148

\bibitem[{{Fesen} {et~al.}(1983){Fesen}, {Gull}, \& {Ketelsen}}]{fgk83}
{Fesen}, R.~A., {Gull}, T.~R., \& {Ketelsen}, D.~A. 1983, \apjs, 51, 337

\bibitem[{{Gao} {et~al.}(2011){Gao}, {Han}, {Reich}, {Reich}, {Sun}, \&
  {Xiao}}]{ghr+11}
{Gao}, X.~Y., {Han}, J.~L., {Reich}, W., {et~al.} 2011, \aap, 529, A159

\bibitem[{{Gao} {et~al.}(2010){Gao}, {Reich}, {Han}, {Sun}, {Wielebinski},
  {Shi}, {Xiao}, {Reich}, {F{\"u}rst}, {Chen}, \& {Ma}}]{grh+10}
{Gao}, X.~Y., {Reich}, W., {Han}, J.~L., {et~al.} 2010, \aap, 515, A64

\bibitem[{{Grosdidier} {et~al.}(2001){Grosdidier}, {Acker}, \&
  {Moffat}}]{gam01}
{Grosdidier}, Y., {Acker}, A., \& {Moffat}, A.~F.~J. 2001, \aap, 370, 513

\bibitem[{{Haffner} {et~al.}(2003){Haffner}, {Reynolds}, {Tufte}, {Madsen},
  {Jaehnig}, \& {Percival}}]{hrt+03}
{Haffner}, L.~M., {Reynolds}, R.~J., {Tufte}, S.~L., {et~al.} 2003, \apjs, 149,
  405

\bibitem[{{Harris} \& {Roberts}(1960)}]{hr60}
{Harris}, D.~E., \& {Roberts}, J.~A. 1960, \pasp, 72, 237

\bibitem[{{Haslam}(1974)}]{has74}
{Haslam}, C.~G.~T. 1974, \aaps, 15, 333

\bibitem[{{Haslam} {et~al.}(1982){Haslam}, {Salter}, {Stoffel}, \&
  {Wilson}}]{hssw82}
{Haslam}, C.~G.~T., {Salter}, C.~J., {Stoffel}, H., \& {Wilson}, W.~E. 1982,
  \aaps, 47, 1

\bibitem[{{Landecker} {et~al.}(1982){Landecker}, {Pineault}, {Routledge}, \&
  {Vaneldik}}]{lprv82}
{Landecker}, T.~L., {Pineault}, S., {Routledge}, D., \& {Vaneldik}, J.~F. 1982,
  \apjl, 261, L41

\bibitem[{{Mavromatakis} {et~al.}(2000){Mavromatakis}, {Papamastorakis},
  {Paleologou}, \& {Ventura}}]{mppv00}
{Mavromatakis}, F., {Papamastorakis}, J., {Paleologou}, E.~V., \& {Ventura}, J.
  2000, \aap, 353, 371

\bibitem[{{Mitra} {et~al.}(2003){Mitra}, {Wielebinski}, {Kramer}, \&
  {Jessner}}]{mwkj03}
{Mitra}, D., {Wielebinski}, R., {Kramer}, M., \& {Jessner}, A. 2003, \aap, 398,
  993

\bibitem[{{Pazderska} {et~al.}(2009){Pazderska}, {Gawro{\'n}ski}, {Feiler},
  {Birkinshaw}, {Browne}, {Davis}, {Kus}, {Lancaster}, {Lowe}, {Pazderski},
  {Peel}, \& {Wilkinson}}]{pgf+09}
{Pazderska}, B.~M., {Gawro{\'n}ski}, M.~P., {Feiler}, R., {et~al.} 2009, \aap,
  498, 463

\bibitem[{{Phillips}(2005)}]{p05}
{Phillips}, J.~P. 2005, \mnras, 362, 847

\bibitem[{{Pineault} {et~al.}(1993){Pineault}, {Landecker}, {Madore}, \&
  {Gaumont-Guay}}]{plmg93}
{Pineault}, S., {Landecker}, T.~L., {Madore}, B., \& {Gaumont-Guay}, S. 1993,
  \aj, 105, 1060

\bibitem[{{Pineault} {et~al.}(1997){Pineault}, {Landecker}, {Swerdlyk}, \&
  {Reich}}]{plsr97}
{Pineault}, S., {Landecker}, T.~L., {Swerdlyk}, C.~M., \& {Reich}, W. 1997,
  \aap, 324, 1152

\bibitem[{{Reich}(1982)}]{rei82}
{Reich}, W. 1982, \aaps, 48, 219

\bibitem[{{Reich} {et~al.}(2004){Reich}, {F{\"u}rst}, {Reich}, {Uyan{\i}ker},
  {Wielebinski}, \& {Wolleben}}]{rei04}
{Reich}, W., {F{\"u}rst}, E., {Reich}, P., {et~al.} 2004, in The Magnetized
  Interstellar Medium, ed. {B.~Uyan{\i}ker, W.~Reich, \& R.~Wielebinski},
  45--50

\bibitem[{{Rengelink} {et~al.}(1997){Rengelink}, {Tang}, {de Bruyn}, {Miley},
  {Bremer}, {Roettgering}, \& {Bremer}}]{rtb+97}
{Rengelink}, R.~B., {Tang}, Y., {de Bruyn}, A.~G., {et~al.} 1997, \aaps, 124,
  259

\bibitem[{{Seward} {et~al.}(1995){Seward}, {Schmidt}, \& {Slane}}]{sss95}
{Seward}, F.~D., {Schmidt}, B., \& {Slane}, P. 1995, \apj, 453, 284

\bibitem[{{Shi} {et~al.}(2008){Shi}, {Han}, {Gao}, {Sun}, {Xiao}, {Reich}, \&
  {Reich}}]{shg+08}
{Shi}, W.~B., {Han}, J.~L., {Gao}, X.~Y., {et~al.} 2008, \aap, 487, 601

\bibitem[{{Sieber} {et~al.}(1979){Sieber}, {Haslam}, \& {Salter}}]{shs79}
{Sieber}, W., {Haslam}, C.~G.~T., \& {Salter}, C.~J. 1979, \aap, 74, 361

\bibitem[{{Sieber} {et~al.}(1981){Sieber}, {Salter}, \& {Mayer}}]{ssm81}
{Sieber}, W., {Salter}, C.~J., \& {Mayer}, C.~J. 1981, \aap, 103, 393

\bibitem[{{Simonetti}(1992)}]{sim92}
{Simonetti}, J.~H. 1992, \apj, 386, 170

\bibitem[{{Si{\'o}dmiak} \& {Tylenda}(2001)}]{st01}
{Si{\'o}dmiak}, N., \& {Tylenda}, R. 2001, \aap, 373, 1032

\bibitem[{{Slane} {et~al.}(1997){Slane}, {Seward}, {Bandiera}, {Torii}, \&
  {Tsunemi}}]{ssb+97}
{Slane}, P., {Seward}, F.~D., {Bandiera}, R., {Torii}, K., \& {Tsunemi}, H.
  1997, \apj, 485, 221

\bibitem[{{Sofue} \& {Reich}(1979)}]{sr79}
{Sofue}, Y., \& {Reich}, W. 1979, \aaps, 38, 251

\bibitem[{{Sokoloff} {et~al.}(1998){Sokoloff}, {Bykov}, {Shukurov},
  {Berkhuijsen}, {Beck}, \& {Poezd}}]{sbs+98}
{Sokoloff}, D.~D., {Bykov}, A.~A., {Shukurov}, A., {et~al.} 1998, \mnras, 299,
  189

\bibitem[{{Stil} {et~al.}(2011){Stil}, {Taylor}, \& {Sunstrum}}]{sts11}
{Stil}, J.~M., {Taylor}, A.~R., \& {Sunstrum}, C. 2011, \apj, 726, 4

\bibitem[{{Sun} \& {Reich}(2010)}]{sr10}
{Sun}, X., \& {Reich}, W. 2010, RAA, 10, 1287

\bibitem[{{Sun} {et~al.}(2007){Sun}, {Han}, {Reich}, {Reich}, {Shi},
  {Wielebinski}, \& {F{\"u}rst}}]{shr+07}
{Sun}, X.~H., {Han}, J.~L., {Reich}, W., {et~al.} 2007, \aap, 463, 993

\bibitem[{{Sun} \& {Reich}(2009)}]{sr09}
{Sun}, X.~H., \& {Reich}, W. 2009, \aap, 507, 1087

\bibitem[{{Sun} {et~al.}(2006){Sun}, {Reich}, {Han}, {Reich}, \&
  {Wielebinski}}]{srh+06}
{Sun}, X.~H., {Reich}, W., {Han}, J.~L., {Reich}, P., \& {Wielebinski}, R.
  2006, \aap, 447, 937

\bibitem[{{Sun} {et~al.}(2011){Sun}, {Reich}, {Han}, {Reich}, {Wielebinski},
  {Wang}, \& {M{\"u}ller}}]{srh+11}
{Sun}, X.~H., {Reich}, W., {Han}, J.~L., {et~al.} 2011, \aap, 527, A74

\bibitem[{{Sun} {et~al.}(2008){Sun}, {Reich}, {Waelkens}, \&
  {En{\ss}lin}}]{srwe08}
{Sun}, X.~H., {Reich}, W., {Waelkens}, A., \& {En{\ss}lin}, T.~A. 2008, \aap,
  477, 573

\bibitem[{{Taylor} {et~al.}(2009){Taylor}, {Stil}, \& {Sunstrum}}]{tss09}
{Taylor}, A.~R., {Stil}, J.~M., \& {Sunstrum}, C. 2009, \apj, 702, 1230

\bibitem[{{Turtle} {et~al.}(1962){Turtle}, {Pugh}, {Kenderdine}, \&
  {Pauliny-Toth}}]{tpkp62}
{Turtle}, A.~J., {Pugh}, J.~F., {Kenderdine}, S., \& {Pauliny-Toth}, I.~I.~K.
  1962, \mnras, 124, 297

\bibitem[{{Umana} {et~al.}(2008){Umana}, {Leto}, {Trigilio}, {Buemi},
  {Manzitto}, {Toscano}, {Dolei}, \& {Cerrigone}}]{ult+08}
{Umana}, G., {Leto}, P., {Trigilio}, C., {et~al.} 2008, \aap, 482, 529

\bibitem[{{Uyan{\i}ker} {et~al.}(1998){Uyan{\i}ker}, {F\"urst}, {Reich},
  {Reich}, \& {Wielebinski}}]{ufr+98}
{Uyan{\i}ker}, B., {F\"urst}, E., {Reich}, W., {Reich}, P., \& {Wielebinski},
  R. 1998, \aaps, 132, 401

\bibitem[{{Wardle} \& {Kronberg}(1974)}]{wk74}
{Wardle}, J.~F.~C., \& {Kronberg}, P.~P. 1974, \apj, 194, 249

\bibitem[{{Xiao} {et~al.}(2008){Xiao}, {F{\"u}rst}, {Reich}, \& {Han}}]{xfrh08}
{Xiao}, L., {F{\"u}rst}, E., {Reich}, W., \& {Han}, J.~L. 2008, \aap, 482, 783

\bibitem[{{Xiao} {et~al.}(2011){Xiao}, {Han}, {Reich}, {Sun}, {Wielebinski},
  {Reich}, {Shi}, \& {Lochner}}]{xhr+11}
{Xiao}, L., {Han}, J.~L., {Reich}, W., {et~al.} 2011, \aap, 529, A15

\bibitem[{{Xiao} {et~al.}(2009){Xiao}, {Reich}, {F{\"u}rst}, \& {Han}}]{xrfh09}
{Xiao}, L., {Reich}, W., {F{\"u}rst}, E., \& {Han}, J.~L. 2009, \aap, 503, 827

\bibitem[{{Xu} {et~al.}(2007){Xu}, {Han}, {Sun}, {Reich}, {Xiao}, {Reich}, \&
  {Wielebinski}}]{xhs+07}
{Xu}, J.~W., {Han}, J.~L., {Sun}, X.~H., {et~al.} 2007, \aap, 470, 969

\bibitem[{{Zhang} {et~al.}(2003){Zhang}, {Reich}, {Reich}, \&
  {Wielebinski}}]{zrrw03}
{Zhang}, X., {Reich}, W., {Reich}, P., \& {Wielebinski}, R. 2003, ChJAA, 3, 347

\end{thebibliography}
\appendix
\section{NGC~40}\label{sect_ngc40}
NGC~40 (${\rm RA}=0^{\rm h}13^{\rm m}2^{\rm s}$, 
${\rm Dec}=72\degr30\arcmin31\arcsec$) is a well-known planetary nebula where its central star shows the 
Wolf-Rayet phenomenon \citep{gam01}. Its distance is very uncertain and 
estimates range 
from 0.5~kpc to 3.5~kpc \citep{p05}. From the polarisation images at 
$\lambda$6~cm (Fig.~\ref{cta16cm}), $\lambda$11~cm (Fig.~\ref{cta111cm}), and 
$\lambda$21~cm (Fig.~\ref{cta121cm}), it is evident that NGC~40 does not have 
an effect on polarisation distribution and cannot cause the low-polarisation 
region, which divides CTA~1 into an eastern and a southern shell. Therefore its 
distance should be larger than that of CTA~1 unless it causes no Faraday 
rotation.

\begin{figure}[!htbp]
\resizebox{0.5\textwidth}{!}{\includegraphics[angle=-90]{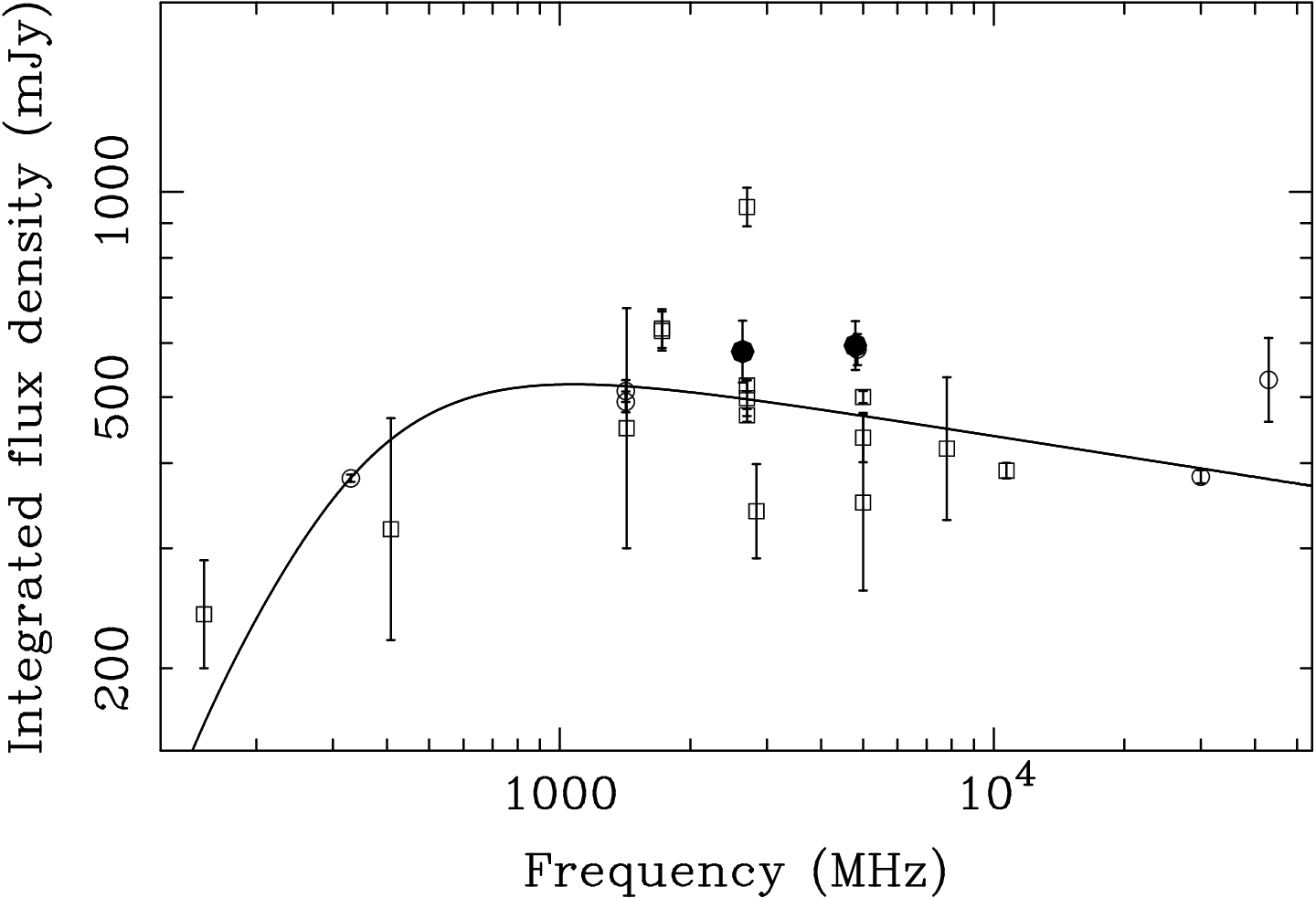}}
\caption{Spectrum of integrated flux densities of NGC~40. The integrated flux 
densities collected by \citet{ssm81} are shown by squares. More flux densities 
were reported at 330~MHz \citep{rtb+97}, 1420~MHz \citep{ccg+98,plmg93}, 
4.85~GHz \citep{bwe91}, 30~GHz \citep{pgf+09}, and 43~GHz \citep{ult+08} and 
shown by open circles. The new $\lambda6$~cm and $\lambda$11~cm measurements 
are marked as filled circles.}
\label{ngc40}
\end{figure}

The integrated flux density of NGC 40 was measured to be $595\pm47$~mJy at 
$\lambda$6~cm and $583\pm58$~mJy at $\lambda$11~cm. Besides the flux density 
measurements collected by \citet{ssm81} we found some new ones from literature 
and plotted both datasets in Fig.~\ref{ngc40}. The two-component radio 
emission model of planetary nebulae by \citet{st01} was used to fit the 
spectrum, which yields an opacity of $\tau=0.0054(\nu/5\,{\rm GHz})^{-2.1}$.

\end{document}